# Investigations on the optical properties and X-ray emission in compact radio AGN

Mai Liao[1,2] | Minfeng Gu[3]

[1] CAS Key Laboratory for Researches in Galaxies and Cosmology, University of Science and Technology of China, Chinese Academy of Sciences, Hefei, Anhui 230026, Peoples Republic of China

[2] School of Astronomy and Space Science, University of Science and Technology of China, Chinese Academy of Sciences, Hefei, Anhui 230026, Peoples Republic of China

[3] Key Laboratory for Research in Galaxies and Cosmology, Shanghai Astronomical Observatory, Chinese Academy of Sciences, 80 Nandan Road, Shanghai 200030, China

**Correspondence**
Mai Liao, Minfeng Gu Email: liaomai@ustc.edu.cn, gumf@shao.ac.cn

Compact radio active galactic nuclei (compact radio AGN) are compact ($\leq 20$ kpc), powerful radio sources. Currently, the preferred scenario is that they are at the early stage of AGN evolution. At present, the research of compact radio AGN mainly focuses on the radio band, other bands have not been extensively studied. We present the systemic optical properties and X-ray emission studies for compact radio AGN, to investigate the accretion properties, AGN evolution and their X-ray origin. We find that compact radio AGN have various accretion modes indicated by the accretion rate analysis. In the radio power–linear size diagram they generally follow the evolutionary trend towards large-scale radio galaxies with increasing linear size and decreasing accretion rate. Their hard X-ray emission may be from jet based on the radio/X-ray relation and fundamental plane of black hole activity.

## 1 | INTRODUCTION

Comapct radio active galactic nuclei (compact radio AGN) are a subclass of AGN, with an intrinsically compact radio morphology ($\leq 20$ kpc) and a peaked radio spectrum, consisting of gigahertz peaked spectrum (GPS) radio sources, compact steep spectrum (CSS) radio sources, and high-frequency peaker (HFP) radio sources based on the convex frequency (see review by O'Dea & Saikia, 2021). They are usually believed to be at an early stage of AGN evolution and will evolve into large-scale radio sources, known as Fanaroff–Riley II/FR I. The youth scenario is strongly supported by the measurements of dynamical and spectral age of $\sim 10^3 - 10^5$ yr (e.g., An & Baan, 2012; Murgia, 2003).

At present, there are only a few studies of optical properties using various samples of compact radio AGN. Wu (2009) and Son et al. (2012) investigated the optical properties of compact radio AGN by using the samples with a few tens sources. Wu (2009) found that most compact radio AGN have relatively high Eddington ratios, which are similar to those of narrow-line Seyfert 1 galaxies (NLS1s) which are a special class of AGN and usually thought to be compact with relatively small black hole masses and high accretion rates (see review by Komossa, 2008). But Son et al. (2012) found accretion activity presented a diversity in compact radio AGN. Notwithstanding these two works give us clues on the accretion process in compact radio AGN, the source number of their sample is rather limited. A larger sample is needed to further systematically study their optical properties, including accretion rate calculated from the optical spectra (see §2.2) and AGN evolution by combining radio and optical properties (see §2.3).

Due to the compactness in compact radio AGN and the resolution limits of present X-ray telescopes, the origin of the X-ray emission is still not clear. Their X-ray emission can be mainly, from the disc–corona system based on observational statistic analysis (e.g., Siemiginowska, LaMassa, Aldcroft, Bechtold, & Elvis, 2008), or either from jets for quasars or radio lobes for galaxies by SED fitting (Migliori, Siemiginowska, & Celotti, 2012; Ostorero et al., 2010). The radio/X-ray luminosity ($L_{\rm R} \propto L_{\rm X}^b$) relation and the fundamental plane (FP) of black hole activity ($\log L_{\rm R} = \xi_{\rm RX} \log L_{\rm X} + \xi_{\rm RM} \log M_{\rm BH} + {\rm constant}$) can be a useful method to study the mechanism of X-ray emission in AGN. Because the radio/X-ray relation with slope b and $\xi_{RX}$ of 0.6−0.7 and 1.4, respectively, could be explained by different accretion modes of radiatively inefficient and efficient accretion as shown in previous various studies (e.g., Merloni, Heinz, & di Matteo, 2003). Fan & Bai (2016) presented the first systematic study of the radio/X-ray relationship and FP for compact radio AGN, and show the accretion-related X-ray emission with b and $\xi_{RX}$ ∼ 0.6. However, higher-resolution radio data (VLBI, VLA, and/or FIRST) compared with their used low resolution radio data of NVSS (45 ″), to exclude the extended radio emission, is crucial to explore the origin of X-ray emission in compact radio AGN.

Therefore, in order to systematically study the accretion mode, and AGN evolution in compact radio AGN, as well as their X-ray emission origin, here we show the optical properties and X-ray emission studies by building the largest optical and X-ray sample at present.

## 2 | OPTICAL PROPERTIES

### 2.1 | Sample and spectroscopic analysis

We have compiled a large radio-selected sample of 468 compact radio AGN, which we regard as a parent sample, by combining various available radio-selected samples in the literature (see more details in Liao & Gu, 2020). The SDSS spectroscopic counterparts were searched within 2 ″ from the NED position for all the objects in parent sample, resulting in our optical sample, consisting of 126 sources.

The spectroscopic analysis for SDSS spectra include two steps that are continnum subtraction and emission lines fitting. We used different models to deal with the continnum subtraction for galaxy-type and quasar-type spectra (see more details in Liao & Gu, 2020). The emission lines were then fitted to obtain their width and flux density by using Gaussian models.

### 2.2 | Accretion mode

We estimated black hole mass ($M_{BH}$) using various empirical relations utilizing the line width and luminosity of broad lines, or the stellar velocity dispersions ($\sigma_*$), or [OIII] line width which was used as surrogate of $\sigma_*$ (see more details in Liao & Gu, 2020). The Eddington ratio $R_{edd} = L_{bol}/L_{Edd}$ was calculated with the bolometric luminosity $L_{bol}$ estimated from the luminosities of broad lines or narrow [OIII] line, and the Eddington luminosity $L_{edd} = 1.38 \times 10^{38} M_{BH}/M_\odot$. The distribution of Eddington ratios for 102 compact radio AGN covers a broader range from $10^{-4.93}$ to $10^{0.37}$ with quasars having significantly higher $R_{edd}$ than galaxies (see details in Fig. 1 ). There exists two accretion modes in our sample if adopting 0.01 of $R_{edd}$ as dividing value between the optically thin advection-dominated accretion flow and optically thick standard accretion disc (Narayan & Yi, 1995), indicating that our compact sources with evolving jet are not only associated with the high-accreted system (Son et al., 2012).

### 2.3 | AGN evolution

The distributions between linear size (LS) and $M_{BH}$, and $R_{edd}$ are plotted in Fig. 1 with showing the comparisons between compact radio AGN and large-scale radio galaxies from Hu, Cao, Chen, & You (2016). Clearly, the spread in $M_{BH}$ for our compact radio sources is much larger than large-scale radio galaxies. Most compact radio AGN (∼ 87%) in Fig 1 . have comparable $M_{BH}$ masses to large-scale radio galaxies (> $10^8 M_\odot$), which may be caused by multiple accretion activity, however the other objects have relatively lower black hole masses. No correlation between $R_{edd}$ and LS, but a trend of decreasing $R_{edd}$ with increasing LS when combined our sources and large-scale radio sources was found. The mean value of $R_{edd}$ in Fig. 1 , of our compact sources is $10^{-2.26}$, while of large-scale radio sources is $10^{-3.05}$.

We further involved $R_{edd}$ ratio into radio luminosity-LS diagram, plotted in Fig. 2 to investigate $R_{edd}$ changing with LS for different radio luminosity sources. From the radio perspective, our sources generally follow the expected evolutionary tracks towards large-scale sources as the smallest high (low) luminosity CSO/GPS/HFP objects to medium-sized high (low) luminosity CSSs to large-scale high (low) luminosity FRII/FRI sources. The trend of decreasing $R_{edd}$ with increasing LS for high/low luminosity radio sources ($L_{5GHz} = 10^{26}$ W/Hz as a boundary) is also evident. The median (log $R_{edd}$) are -1.44 and -2.52 for high luminosity compact radio sources and large-scale FR II radio galaxies, respectively. And for low luminosity compact radio sources and large-scale FR I radio galaxies, the median (log $R_{edd}$) are -3.01 and -4.23, respectively. These results strongly indicate that the compact radio AGN will eventually evolve into large-scale radio sources in general. But see the discussion on the compact radio sources with low luminosity and low accretion rate in Liao & Gu (2020), which suggest that some are short-lived sources and can't grow into large-scale ones.

## 3 | X-RAY EMISSION

### 3.1 | Sample and data

Our X-ray sample consists of 91 sources, which was built by cross-matching our parent sample with Chandra and XMM-Newton X-ray archives, and including additional two CSS radio sources of 3C 305 and 4C 13.66. After excluding six objects that are too faint and either lensed or significant in X-ray spectral variability (see more details in Liao, Gu, Zhou, & Chen, 2020), the X-ray data is thus available in 85 sources.

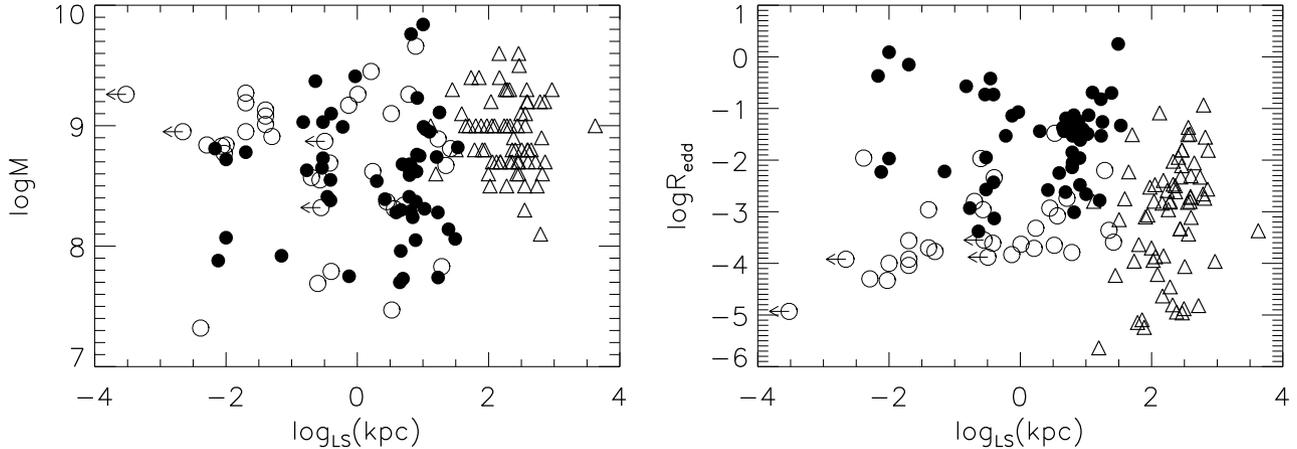

**FIGURE 1** $M_{BH}$ versus LS (*left*) and $R_{edd}$ vs. LS (*right*). The filled and open circles are compact radio AGN classified as quasar and galaxy, respectively. The triangles are large-scale radio galaxies in Hu et al. (2016). The arrows represent the upper limits on LS.

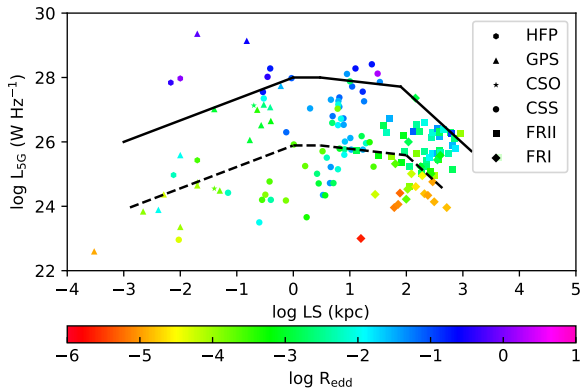

**FIGURE 2** $L_{5GHz}$ vs. LS with $R_{edd}$. Here compact symmetric objects (CSOs) with morphologically classified are also included in our sample, which share similar properties as GPS radio galaxies. The color bar indicate the value of $R_{edd}$. The black solid (dashed lines) are the expected evolutionary tracks based on parametric modeling for the high-luminosity (low-luminosity radio sources) in An & Baan (2012).

Their X-ray information of flux density at $2-10$ keV and photon index ($\Gamma$) that are required to study the X-ray emission, were collecting from literature, or obtained from the data reduction by ourselves. The published VLBI and VLA data at 5 GHz is only for radio core from NED and the literature, with the aim to minimize the contamination from extended radio emission was collected. The FIRST 1.4 GHz flux (Helfand, White, & Becker, 2015) is also included when available as following comparative study. And $M_{BH}$ and $R_{edd}$ were directly taken from our optical studies.

### 3.2 | Radio/X-ray relation and Fundamental plane

We firstly investigated the radio/X-ray luminosity relation for the whole X-ray sample. We found strong correlations between the X-ray luminosity $L_X$ at $2-10$ keV and the radio luminosity $L_R$ at 5 GHz of VLA, FIRST, and VLBI, implying that both pc- and kpc-scale radio emission have tight relation with X-ray emission. The OLS bisector (Isobe, Feigelson, Akritas, & Babu, 1990) fits to the correlations between $L_X$ and $L_R$ give that approximately linear relations between log $L_X$ and log $L_R$ with the slope b equal about 1 (see more details in Liao et al., 2020).

Since the the slope of radio/X-ray emission relation can be quite different between radiative efficient and inefficient accretion flows (see introduction), we used a dividing value of $R_{edd} = 0.001$ (Qiao & Liu, 2015) to distinguish accretion mode for 34 objects which have both BH mass and Eddington ratio estimations, resulting 26 sources with radiative efficient flow (i.e., $R_{edd} > 10^{-3}$)[1]. The source number reduces to 13 (12) when VLBI radio-core (FIRST[2]) data is further required. The multiple linear regression of Bayesian approach (Kelly, 2007) was applied to investigate FP for high accreted compact radio AGN as did in Fan & Bai (2016). Our best fitting results show that $\xi_{RX}$ in FP equal $0.99^{+0.05}_{-0.05}$ and $1.08^{+0.07}_{-0.07}$ for VLBI data (13 sources) and FIRST data (12 sources) with $R_{edd} > 10^{-3}$, respectively. And the radio /X-ray relation b equals to $1.02^{+0.14}_{-0.14}$ ($1.15^{+0.23}_{-0.23}$) for VLBI (FIRST) within FP sources.

---

[1]Here, we adopted a different dividing value compared with the one in section 2.2 due to there still has no clear boundary between two accretion modes and enable us to work FP using limited sources with available $M_{BH}$, $R_{edd}$, $L_X$ and $L_X$ simultaneously.

[2]The FIRST data can better represent kiloparsec radio emission than VLA data owing to the consistent resolution.

## 3.3 | X-ray origin

Both b and $\xi_{RX} \sim 1$ for the high-accreting sources in our sample deviate from the theoretical models predictions, that the radio/X-ray slope b and $\xi_{RX}$ would be $\sim 1.4$ when X-ray emission is from efficient accretion flow (Merloni et al., 2003), indicating the X-ray emission in our high-accreting sources is not dominated by accretion process, and the contribution from the jet might be important as they are strong radio emitters. We failed to find any significant correlations between $\Gamma$ and $R_{edd}$ for our sources with VLBI core and FIRST detections, where the $\Gamma$-$R_{edd}$ relation has been broadly found in radio-quiet AGN with the clue on the X-ray emission from disk-corona system (e.g., Brandt & Alexander, 2015). Moreover, the average value of $\Gamma$ of the sources included in the FP is 1.64, which is similar to that of flat spectrum radio quasars(FSRQs, 1.65±0.04, Donato, Ghisellini, Tagliaferri, & Fossati, 2001) and is flatter than normal radio-quiet quasars (RQQs) ($\Gamma \sim 1.89$, Körding, Falcke, & Corbel, 2006). All the results suggest that the hard X-ray emission of our high-accreted sources could be different from RQQs and may be from jet like FSRQs (Chen, 2018).

Compared with Fan & Bai (2016) of b and $\xi_{RX} \sim 0.6$, which support the accretion-related X-ray emission, the discrepancy between both works may be caused by using the different radio data that has different resolution because both are based on the comparable sample size and with high accretion rate of $R_{edd} > 10^{-3}$. After carefully checking the data that Fan & Bai (2016) used, we found that their $L_X$, all based on the original data from Siemiginowska et al. (2008) and Tengstrand et al. (2009), were incorrectly overestimated (private communication with authors). After correcting $L_X$, we found their re-analyzed FP relation $\xi_{RX} = 0.92 \pm 0.05$ is consistent with our result (0.99±0.05) within errors for their sources with $R_{edd} > 10^{-3}$.

## 4 | SUMMARY

We constructed the largest optical and X-ray sample for compact radio AGN, and systematically study their optical properties and the origin of X-ray emission. We find that they have mixed accretion modes, and they generally follow the evolutionary trend towards large-scale radio galaxies with increasing linear size and decreasing accretion rate in the radio power-linear size diagram. The X-ray emission of high-accreting compact radio AGN may be from jets.

## ACKNOWLEDGMENTS

This work is supported by the National Science Foundation of China (Grants No. 11873073, 11773056, U1831138) and the China Postdoctoral Science Foundation (Grant No. 2021M693089).

## 5 | AUTHOR BIOGRAPHY

**Mai Liao** obtained her PhD in Shanghai Astronomical Observatory of China, and is currently a postdoctoral researcher of University of Science and Technology of China. Her interests include AGN multi-band emission, AGN outflows, and AGN variability.